\documentstyle[12pt]{article}
\begin{document}
\newcommand{\be}{\begin{equation}}
\newcommand{\ee}{\end{equation}}
\newcommand{\ea}{\end{array}}
\newcommand{\dis}{\displaystyle}

\rightline{FTUV/94-29}
\leftline{Submited to ICHEP94}

\begin{center}
{\LARGE{\bf Neutrino mass and magnetic moment
 from neutrino-electron scattering.}}
\end{center}

\vspace{2cm}

\begin{center}
{\large{\bf J. Bernab\'{e}u, S.M. Bilenky, F.J. Botella,
 J.A. Pe\~{n}arrocha and J. Segura}}

\vspace{1cm}

{\it Departament de F\'{\i}sica Te\'{o}rica \\ Universitat de Val\`{e}ncia \\
and \\ IFIC, Centre Mixt Univ. Valencia-CSIC \\ E-46100 Burjassot,
Spain}
\end{center}

\vspace{2cm}

\baselineskip 0.7cm

\begin{center}
{\bf Abstract}
\end{center}
{\noindent
  We study both the elastic ($\nu e \rightarrow \nu e$) and the radiative
 process ($\nu e \rightarrow \nu e \gamma$) and discuss how these
 processes can shed light on some current topics in neutrino physics such
 as a neutrino magnetic moment and neutrino oscillations. The radiative
 process allows to reach low values of $Q^2$ without the need to operate
 at very small energies of recoil electrons, a favourable scenario to search
 for a neutrino magnetic moment. The elastic cross section contains a dynamical
 zero at $E_{\nu}=m/(4 sin^2\theta_{W})$ and forward electrons for the
 electron antineutrino channel, which is reachable at reactor facilities
 and accessible after the convolution with the antineutrino spectrum. The
 implication for lepton flavour changing transitions in that energy region
 searched for in neutrino oscillation experiments, which combine disappearance
 and appearance rates, is discussed.}

\newpage

\section{Introduction.}

 The neutrino-electron process
 plays a crucial role in the study of the standard model
of electroweak interactions, as well as in searching for effects beyond the
standard model.
 It gives
relevant information about possible deviations from S.M.
as, for instance, the possible existence of a large neutrino magnetic moment:
the laboratory bound  on the neutrino magnetic moment
($\mu_{\nu}<
2.4\small{x}10^{-10}$) has been set \cite{Vyd} with $\bar{\nu}_{e} e^{-}
\rightarrow \bar{\nu}_{e} e^{-}$  in reactor experiments and
several new proposals \cite{Bro} plan
 to study the $\bar{\nu}_{e}$ magnetic moment at the level of $2\small{x}
10^{-11}$
Bohr magnetons .

The differential cross
section for $\bar{\nu}_{i} e^{-}
\rightarrow \bar{\nu}_{i} e^{-}$ including the neutrino magnetic moment
 contribution \cite{Vog} and neglecting
 neutrino mass
    is given by

\begin{equation}
\frac{\displaystyle{d\sigma_{\bar{\nu}_{e}}}}{\displaystyle{dT}}=
\frac{\displaystyle{G^{2}m}}{\displaystyle{2\pi}}\left[ (g^{i}_{R})^{2}
+(g^{i}_{L})^{2}\left(1-\frac{\displaystyle{T}}
{\displaystyle{E_{\nu}}}\right)^{2}
-g^{i}_{L}g^{i}_{R}\frac{\displaystyle{mT}}
{\displaystyle{E_{\nu}^{2}}}\right]
+\frac{\displaystyle{\pi \alpha^{2} }}{\displaystyle{m^{2}}}
\left(\frac{\displaystyle{\mu_{\nu}}}{\displaystyle{\mu_{B}}}\right)^2
\frac{\displaystyle{(1-T/E_{\nu})}}{\displaystyle{T}}
\end{equation}
\

\noindent
where G is the Fermi coupling constant, $\alpha$ the fine structure
constant, $\mu_{\nu}$ the neutrino magnetic moment, $\mu_{B}$ the Bohr
 Magneton,
$m$ the electron mass, T the   recoil kinetic
energy of the electron and
$E_{\nu}$ the antineutrino incident energy. In terms of
 $sin^{2}\theta_{W}$
the chiral couplings $g^{i}_{L}$ and $g^{i}_{R}$ ($i=e,\mu,\tau$)
can be written as

\begin{equation}
\begin{array}{cc}
g^{i}_{L}=-1+2sin^{2}\theta_{W}+2\delta_{ie};
 & g^{i}_{R}=2sin^{2}\theta_{W} \,\, .
\end{array}
\end{equation}
\

\noindent
( for neutrinos one should exchange
$g_{L}$ by
 $g_{R}$ ).

 In the laboratory experiments on neutrino magnetic moment,
 the sensitivity to $\mu_{\nu}$ is connected with
the fact that at low enough values of $Q^{2}=2mT$ the contribution of the
electromagnetic amplitude to the cross section of the process becomes
comparable to the contribution of the weak amplitude.
This is the case for $Q^{2} \sim MeV^{2}$ at values
$\mu_{\nu}\simeq (10^{-10}$, $10^{-11}) \mu_{B}$.
 The penetration in the region of such small
$Q^{2}$ requires, however, to measure small  energies of recoil
electrons ($T\leq MeV$).

In section 2 we discuss the use of the radiative process to extract
 a neutrino magnetic moment. Section 3 presents the dynamical zero in elastic
 scattering, whereas section 4 shows a novel
 approach to neutrino oscillations based on this zero.

\section{Neutrino magnetic moment and the radiative process.}

 Now let us consider the process
$\nu (\bar{\nu}) + e \rightarrow \nu (\bar{\nu}) + e + \gamma$ .
 Even if it
 has an additional power of
$\alpha$ in the cross section relative to the elastic case, the restriction
to low recoil energies in order to reach down low values of $Q^{2}$ is a priori
not necessary. The limit $Q^{2} = 0$ at fixed values of the
recoil energies can be reached for the
maximal opening angle between electron and photon in the final state
($\theta_{e\gamma}$). Whatever
 the
experimental limit on the total recoil energies $\nu$ could be, this
process is able to lead to lower values of $Q^{2}$ than the elastic one,
as shown by the ratio $x = Q^{2} /(2 m \nu)$ varying from 1 to 0 ($\nu=
T+E_{\gamma}$, being $E_{\gamma}$ the photon energy).

 With this motivation we have calculated \cite{Ber} the triple differential
cross section for the process

\begin{equation}
\nu (l) + e (p) \rightarrow \nu (l') + e (p') + \gamma (k)
\end{equation}

\noindent
in terms of the three dimensionless variables

\be
\begin{array}{c}
x = Q^{2} /(2m \nu) \,\,\,
 y = \frac{\nu}{E_{\nu}}\,\,\,
\omega = \frac{E_{\gamma}}{E_{\nu}}
\ea
\ee

\noindent
both for the weak and the magnetic contribution, which add incoherently
 due to the opposite final neutrino helicities induced by each of these
 two interactions for massless neutrinos.
 For fixed $x$ and $y$, the $\omega$-integration
 in the cross section can also be performed in an analytic way.

We are interested in the behaviour of both the weak and the electromagnetic
cross sections at low $Q^{2}$, with a view to enhance the second contribution
with respect to the first one.
 First we consider, at $y$, $\omega$ fixed, the
expansion around $x \rightarrow 0$. The weak cross section is

\be
\begin{array}{rl}
\frac{\displaystyle{d \sigma_{W}}}{\displaystyle{d x d y d \omega}}_{x<<1}
\simeq &
 \frac{\displaystyle{G^{2} m^{2}}}{\displaystyle{\pi^{2}}}\left.
\alpha \frac{\displaystyle{1}}{\displaystyle{y^{3} \omega}}
\right\{ W (y, \omega) g^{2}_{A}\\
\\
&  \left.+ \frac{\displaystyle{E_{\nu} x y}}
{\displaystyle{2m}} [V (y, \omega,\frac{m}{E_{\nu}})
g^{2}_{V} + A (y, \omega,\frac{m}{E_{\nu}})
g^{2}_{A} +  I (y, \omega) g_{V} g_{A} ] \right\}
\end{array}
\ee

\noindent
where

\be
\begin{array}{ll}
W (y, \omega)  = & (1 - y) ( y - \omega)^{2}
\end{array}
\ee

\noindent
$V(y,\omega,\frac{m}{E_{\nu}})$,
$A(y,\omega,\frac{m}{E_{\nu}})$ and $I(y,\omega)$ are well behaved functions
 and all the explicit $E_{\nu}$ dependences come in powers of $m/E_{\nu}$.

The couplings are

\be
g_{V} = \frac{g_{L} + g_{R}}{2}, \, g_{A} = \frac{g_{L} - g_{R}}{2}
\ee

\noindent
in terms of the chiral couplings of Eq. (2) ( going form $\nu$ to $\bar{\nu}$
 one should change the $g_{A}$ sign).

There are interesting features associated with this result. At $x = 0$ the
only survival term in the cross section goes like $g^{2}_{A}$.
 It is well known that,
due to CVC, the structure function associated with inelastic excitations
mediated by the vector current goes like $Q^{2}$ at fixed $\nu$.
So only the (PCAC) $g_{A}^{2}$-term \cite{Seh} can survive at $x = 0$.

Nevertheless, $W(y,\omega)$ will be the dominant
 term only in a
very restricted range around $x=0$. So, for example,
this term gives a good approximation
provided $\nu>>m$ (high incoming energies) but within the restricted range
$Q^{2} < <4 m^{2}$. This is so because
the linear term in $x$, in fact, goes as
 $Q^{2}/ 4 m^{2}$.
Furthermore, the $W (y, \omega)$ dependence goes like the square
of the recoil energy of the electron. If $\nu << m$ there are high
 cancellations
in this term, seen for example when one integrates over $\omega$ at fixed $y$.
There is no such cancellation for $V(y,\omega,\frac{m}{E_{\nu}})$,
$A(y,\omega,\frac{m}{E_{\nu}})$ or
 $I(y,\omega)$.
We conclude that the $x = 0$ term is only important at high incoming
energies with $\nu >> m$, but with $Q^{2} << 4m^{2}$.
Our strategy will be just the contrary, i.e., have $\nu < m$ with low
$Q^{2}$, in order to suppress the $x = 0$ $g_{A}^{2}$-term in the weak cross
 section.

The cross section induced by a neutrino magnetic moment $\mu_{\nu} \not=
0$ gives, in the limit $x \rightarrow 0$.

\be
\begin{array}{ll}
\frac{\displaystyle{d \sigma_{M}}}{\displaystyle{d x dy d\omega}}_{ x < < 1}
 \simeq
& \frac{\displaystyle{\alpha^{3}}}{\displaystyle{2 m^{2}}}
 \left( \frac{\displaystyle{\mu_{\nu}}}
{\displaystyle{\mu_{B}}}\right)^{2}
\frac{\displaystyle{1}}{\displaystyle{y^{3} \omega}}
\left\{ M (y,\omega,\frac{m}{E_{\nu}}) +
x N (y, \omega,\frac{m}{E_{\nu}}) \right\}
\end{array}
\ee
\\

\noindent
where

\be
\begin{array}{ll}
M (y, \omega,\frac{m}{E_{\nu}}) =
& (1 - y) [(y^{2} + \omega^{2}) - 2 \frac{\displaystyle{m}}
{\displaystyle{E_{\nu}}} ( y -
\omega) + \frac{\displaystyle{m^{2}}}{\displaystyle{E_{\nu}^{2} y \omega}}
(y - \omega)^{2}]
\end{array}
\ee

\noindent
and $N(y,\omega,\frac{m}{E_{\nu}})$ is a well
 behaved function with the same explicit $E_{\nu}$
 dependences as $V(y,\omega,\frac{m}{E_{\nu}})$ and
$A(y,\omega,\frac{m}{E_{\nu}})$.

The first point to be noticed in Eq. (8) is the absence of the $1/x$
singularity associated with the photon propagator in the
magnetic contribution present in the
elastic scattering cross section. This is again due to the conservation of
the electromagnetic current in the electron vertex, implying a linear
 $Q^{2}$-behaviour of the structure function, at $\nu$ fixed,  for
inelastic excitations. The leading $M (y, \omega)$ term
 is related to the Compton scattering cross section ( like
$V (y, \omega)$ is, and also $A(y,\omega)$ for $E_{\nu}>>m$). In
fact, one can write

\be
\left.\frac{d \sigma_{M}}{d x d y d \omega} \right|_{x = 0} =
\frac{\alpha}{2 \pi} \frac{E_{\nu}}{m} (\frac{\mu_{\nu}}{\mu_{B}})^{2}
(1 - y) \frac{d \sigma^{\gamma \gamma}}{d \omega}
\ee

\noindent
with $\sigma^{\gamma \gamma}$ given by the Klein-Nishina formula identifying
 $y$ with the energy of the incoming photon and $\omega$ with the energy
 of the outgoing photon. Contrary to the behaviour that
we have discussed for $W (y, \omega)$ in the weak cross section, the term
$M( y, \omega)$ is not here suppressed with respect to the linear term
in $x, N (y, \omega)$, so Eq. (10) is a very good approximation to the
magnetic cross section at low energies and low values of $Q^{2}$. Taking
the ratio of cross sections at $Q^{2} = 0$, we have

\be
\begin{array}{rl}
\left.\frac{\displaystyle{d \sigma_{M}}}{\displaystyle{d \sigma_{W}}}
\right|_{x = 0}  = & (\frac{\displaystyle{\mu_{\nu}}}
{\displaystyle{\mu_{B}}})^{2}
\frac{\displaystyle{\pi^{2} \alpha^{2}}}{\displaystyle{G^{2} m^{2}}}
\frac{\displaystyle{1}}{\displaystyle{2 m g^{2}_{A} T}}  \\
\\
& \times  \left\{ \frac{\displaystyle{2 E_{\gamma} (E_{\gamma} + T)
+ T^{2}}}{\displaystyle{m T}} +
\frac{\displaystyle{m T - 2 E_{\gamma} (E_{\gamma} + T)}}
{\displaystyle{E_{\gamma} (E_{\gamma} + T)}}
\right\}
\end{array}
\ee
\\

\noindent
where the global factor in front of the bracket is a typical measure of
this ratio for the elastic scattering process at the same value of $T$.
A glance at
eq. (11) would say that the highest cross section
 ratios are obtained for the hardest
photon limit $E_{\gamma} >> T$, with values higher than the elastic ones at
will. Even more, one would say that higher neutrino energies are favoured
 in order to have hard photons but
 the discussion after eq. (7) should have clarified that a little departure
from $x = 0$ under these conditions is enough to enhance  the
next linear term in $x$ so that the ratio (11) becomes diluted.
To conclude, the strategy to reach low enough $Q^{2}$-values, approaching
$\theta_{max}$ at fixed $(y, \omega)$, works only in a very limited angular
range around $\theta \simeq \theta_{\max}$. Whenever the results are integrated
over a wider region of $\theta$, the
 ratio $d \sigma_{M} / d \sigma_{W}$ will be
diluted, as illustrated in Fig. 1.

We can consider the approach to $Q^{2} \rightarrow 0$ for fixed $x$.
 The vector contribution is in this case not
penalized due to CVC with respect to the axial
 contribution, as it was the case for $x\rightarrow 0$: the structure function
goes like $Q^{2} / \nu$ and the limit $\nu \rightarrow 0$ is not
physically forbidden for our process. It is thus of interest to study the
inclusive cross sections $d \sigma / dx dy$ and explore their
behaviour when $y \rightarrow 0$ at
fixed $x$. We can use the analytic results of the triple differential
 cross sections for the integration in
$\omega$, with the condition $\nu < < m$, and obtain

\begin{equation}
\begin{array}{ll}
\frac{\dis{d^2 \sigma_{W}}}{\dis{d x d \nu}} \simeq & \frac{\dis{4}}{\dis{3}}
 \frac{\dis{G^{2} \alpha}}{\dis{\pi^{2}}}
\frac{\dis{1}}{\dis{1-x}}\nu
\left\{x[ (g_{V}^{2} + g_{A}^{2}) - \frac{\dis{\nu}}{\dis{E_{\nu}}}
(g_{V}^2+g_{A}^2-2xg_{V}g_{A})-\frac{\dis{x}}{\dis{2}}
\frac{\dis{m \nu}}{\dis{E_{\nu}^2}} (g_{V}^2-g_{A}^2)]\right.\\
&\\
&\left.+ \frac{\dis{\nu}}{{m}}[ (\frac{\dis{17}}{\dis{10}}-2)x g_{V}^2
 +\frac{\dis{1}}{\dis{10}}(37x^2-60x+20) g^{2}_{A} ] + O(\nu^2)
\right\}
\end{array}
\ee

\noindent
for the weak cross section, whereas

\begin{equation}
\begin{array}{ll}
\frac{\dis{d^2 \sigma_{M}}}{\dis{dx d \nu}} \simeq &
\frac{ \dis{4 \alpha^{3}}}{\dis{3 m^{3}}} \left(
\frac{\dis{\mu_{\nu}}}{\dis{\mu_{B}}}\right)^{2} \frac{\dis{1}}{\dis{1-x}}
\left\{1 -\frac{\dis{\nu}}{\dis{E_{\nu}}}+\left(\frac{\dis{17}}{\dis{10}}x
-2\right)\frac{\dis{\nu}}{\dis{m}}+O(\nu^2)\right\}
\end{array}
\ee

\noindent
gives the magnetic moment cross section, which is much less sensitive to
low $x$ values.  Note that the $g_{A}^2$ term is the only one
 which survives at $x=0$ with a $\nu^2$ suppression due to the cancellation
 in $W(y,\omega)$.

Fig.2  gives the ratio of the inclusive cross section
$d^2\sigma^{M}/d\nu dx$ over $d^2\sigma^{M}/d\nu dx$ for
electron-antineutrino scattering at \mbox{$E_{\nu} = 1 MeV$}.
This results confirms, as can be guessed from eqs. (12) and (13),
 that the highest sensitivity is obtained for the lowest values of
$\nu$ and, by going down to low values of $x$, the sensitivity is
higher than for the elastic scattering case with $x = 1$. On the contrary,
once $\nu$ is high enough, the  sensitivity is not
improved when lowering the value of $x$. Then we see that,
 although the absolute cross sections are small ( for instance,
$\sigma_{M}/\sigma_{W}=4.4$, $\sigma_{M}=2.7\, 10^{-47} cm^2$ for $\mu_{\nu} =
10^{-10}  \mu_{B}$ integrating over $\nu<0.5 MeV$ ,
 $x<0.5$ )
, the standard model contribution is suppressed in these
circumstances more strongly than in the elastic scattering case, thus giving
 a favourable ratio.
The general features are not highly sensitive to the incoming neutrino energy
within the range of the reactor antineutrino spectrum.

 For $x\rightarrow 1$ and
$\nu \rightarrow 2E_{\nu}^2/
(2E_{\nu}+m)$
one can  see at $E_{\nu} = 1 MeV$ a remarkable feature: the ratio
 $d\sigma^{M}/d\sigma^{W}$
increases rapidly. The limit
 $x\rightarrow 1$ leads
to the infrared behaviour; then the peak-effect should
 also take place for the elastic process. Taking a glance at Eq. (1) one
 concludes that this effect must be a consequence of some cancellation in the
 elastic weak cross section for forward electrons
 \mbox{($T=2E_{\nu}^2/(2E_{\nu}+m)$)}, as we are going to show now.

\section{The dynamical zero in $\bar{\nu}_{e} e^-$ elastic scattering.}

 \hspace{0.5cm}We see from eq. (1) that cancellations may happen in the elastic
 cross section provided $g_{L}g_{R}>0$, as
 is the case for $\bar{\nu}_{e} e^{-}$ scattering. In fact, a complete
 cancellation is going to take place.

  In the LAB frame, the backward neutrino ( forward electron )
cross section for $\nu_{i} e^{-}
\rightarrow \nu_{i} e^{-}$
 can be written as

\begin{equation}
\left(\frac{\displaystyle{d\sigma_{\nu_{i}}}}{\displaystyle{dT}}\right)_{back}
=\frac{\displaystyle{G^{2}m}}{\displaystyle{2\pi}}
\left[ g^{i}_{L}-g^{i}_{R}\frac{\displaystyle{m}}{\displaystyle{
2E_{\nu}+m}}\right]^{2}
\end{equation}
\

\noindent
We see this backward cross section does not cancel
with $g_{L}^{e}$ and $g_{R}^{e}$ satisfying $g_{L}^{e}>g_{R}^{e}>0$,
which is the
case for $\nu_{e}$ as seen from equation (2). On the other hand, for
$\bar{\nu}_{i} e^{-}$ backward elastic scattering we have

\begin{equation}
\left(\frac{\displaystyle{d\sigma_{\bar{\nu}_{i}}}}{\displaystyle{dT}}
\right)_{back}
=\frac{\displaystyle{G^{2}m}}{\displaystyle{2\pi}}
\left[ g^{i}_{R}-g^{i}_{L}\frac{\displaystyle{m}}{\displaystyle{
2E_{\nu}+m}}\right]^{2}
\end{equation}
\

\noindent
which  vanishes for $\bar{\nu}_{e}$ at

\begin{equation}
E_{\nu}=
\frac{\displaystyle{m}}{\displaystyle{4sin^{2}\theta_{W}}}\, \, .
\end{equation}
\

\noindent
Therefore we have found that for the  antineutrino energy $E_{\nu}$ given
by equation (16) and forward electrons ( electrons with maximum
 recoil ) the
differential cross section for $\bar{\nu}_{e}e^{-}\rightarrow \bar{\nu}_{e}
e^{-}$
vanishes exactly at leading order.
This cancellation, depending on the values of $g_{L}$ and $g_{R}$, is a
 dynamical zero \cite{Seg1}.

For $\nu_{\mu}$ and $\bar{\nu}_{\mu}$ elastic scattering ( or
$\nu_{\tau}$ and $\bar{\nu}_{\tau}$ ) the corresponding $g_{L}^{\mu}$,
 $g_{R}^{\mu}$ parameters are such that $g_{L}^{\mu}g_{R}^{\mu}<0$, thus
 preventing the corresponding cross sections from dynamical zeros
for backward neutrinos.

 Studying the conditions that define the potential dynamical zeros
for each  helicity amplitude at lowest order in electroweak
 interactions one obtains all the information
 about dynamical zeros for polarized and
unpolarized differential cross sections.
Let us denote by
 $M_{\lambda ' \lambda }^{\nu_{i}(\bar{\nu}_{i})}$ the helicity amplitudes
 in the LAB frame for
 \mbox{$\nu_{i}(\bar{\nu}_{i}) e^{-}
\rightarrow \nu_{i}(\bar{\nu}_{i}) e^{-}$}, $i=e,\mu,\tau$, being
$\lambda$ and $\lambda '$ the initial and final electron
 helicities respectively (the helicity of the target electron,at rest, is
 referred to the backward direction).

 The conclusions about the dynamical zeros in the helicity
 amplitudes are the following:

{\bf i)} $M_{++}^{\bar{\nu}_{e}}$ shows dynamical zeros given
 in the
energy range

\noindent
 \mbox{$0\leq E_{\nu}\leq m/4sin^{2}\theta_{W}$}.
 The upper value
 corresponds to the phase space point \mbox{$cos\theta =1$}
( where $\theta$ is the
angle in the scattering plane of the final electron
with respect
to the incoming neutrino direction) .
At this end point the
 other
three helicity amplitudes have  kinematical zeros. This is the reason why
this dynamical zero shows up in the unpolarized cross section in the
backward configuration.

{\bf ii)} $M_{-+}^{\bar{\nu}_{\mu},\bar{\nu}_{\tau}}$ show dynamical zeros
 in the whole range of energies $0\leq E_{\nu}<\infty$ .
 In this case the
helicity amplitudes never vanish simultaneously. Then, the dynamical
 zeros will only show up in polarized cross sections.

{\bf iii)} There are no more dynamical zeros for any helicity amplitude
 in the physical region.\\

These results are summarized in Figure 3, where the dynamical zeros are plotted
in the
 plane $(E_{\nu},cos\theta)$ , together with the kinematical zeros.

\section{The dynamical zero and a new kind of neutrino oscillation experiment.}

\hspace{0.5cm}It seems difficult to design a $\bar{\nu_{e}} e^{-}$ experiment
where electron
polarizations are involved. So we shall concentrate in the dynamical zero
 for the unpolarized $\bar{\nu}_{e}-e^-$ elastic cross section.
The fact that the weak backward cross section for
$\bar{\nu_{e}} e^{-}\rightarrow
\bar{\nu_{e}} e^{-}$ vanishes at leading order for $E_{\nu}=
m/(4sin^{2}\theta_{W})$ clearly points out that this kinematical
 configuration must be a good
place to study new physics. Let us stress that backward neutrinos mean forward
electrons, with maximum recoil energy; the electron recoil energy corresponding
 to the dynamical zero  $T\simeq 2m/3$ is in fact within the range of the
 proposed detectors to measure recoil electrons. The neutrino energy for
 which we find the dynamical zero is on the peak of any typical antineutrino
 reactor spectra \cite{Vog,Hah}, being precisely $\bar{\nu}_{e}$
 the flavour which is
 produced copiously in nuclear reactors. We have checked that although
 the dynamical
 zero appears for a given $E_{\nu}$, the convolution of the cross section
 with the antineutrino spectrum still keeps the effect for the planned
 detectors that select neutrino energies by measuring $T$ and the recoil angle
 of the electron. The kinematical region where
 the dynamical zero lies is thus in principle reachable by experiment.

 As a first illustration of the interest of this dynamical zero we shall
concentrate in the possibility of searching for a neutrino magnetic moment
. In Figure 4 we denote by
$(d\sigma_{W}/dT)_
{back}$ the standard contribution in the r.h.s. of eq. (1)
and by $(d\sigma_{M}/dT)_{back}$
the magnetic moment contribution
 , both for $T=T_{max}$.
The solid line represents the boundary where
$(d\sigma_{W}/dT)_
{back}=(d\sigma_{M}/dT)_
{back}$ for $\bar{\nu_{e}} e^{-}\rightarrow
\bar{\nu_{e}} e^{-}$. The regions below the other lines are those for which
$(d\sigma_{M}/dT)_
{back}>(d\sigma_{W}/dT)_
{back}$
 for the rest of neutrino species. It is quite apparent from this
figure that electron antineutrinos with energies around 0.5 MeV give the
possibility of studying low values for neutrino magnetic moment. With other
kind of neutrinos this is only possible by going to much lower values of
neutrino energy.

 The fact that the weak cross section for $\bar{\nu}_{e}$ behaves in such
 a peculiar way in contrast to the other neutrino species suggests a second
 phenomenological implication: measuring neutrino oscillations \cite{Seg2}.

 Let us consider the measurement of the elastic cross section of electron
 antineutrinos, coming from a nuclear reactor, using a detector at some
 distance $x$ from the source. We know that, due to the dynamical zero,
 it is not possible to find forward electrons with $T\simeq 2m/3$ due
 to the $\bar{\nu}_{e}$ $e^-$ interaction. If one of these events is found
 one would conclude that the electron antineutrino has oscillated in the
 way to the detector to another type of neutrino ( $\bar{\nu}_{\mu}$ or
  $\bar{\nu}_{\tau}$ ).

 Then, suppose we have a source of
 electron-antineutrinos $\bar{\nu}_{e}(0)$ ( a
 nuclear reactor for example ) and we measure the differential cross
 section for the process $\bar{\nu}_{e}(x) e^- \rightarrow \bar{\nu}_{e}(x)
 e^-$ at a distance $x$ from the source. If vacuum oscillations take
 place we will have

\begin{equation}
\frac{\displaystyle{d\sigma^{\nu}(E_{\nu},T)}}{\displaystyle{dT}}=
P_{\bar{\nu}_{e}\rightarrow \bar{\nu}_{e}}(x)
\frac{\displaystyle{d\sigma^{\bar{\nu}_{e}}(E_{\nu},T)}}{\displaystyle{dT}}
+\sum_{i=\mu,\tau}P_{\bar{\nu}_{e}\rightarrow \bar{\nu}_{i}}(x)
\frac{\displaystyle{d\sigma^{\bar{\nu}_{i}}(E_{\nu},T)}}{\displaystyle{dT}}
\end{equation}

\noindent
where $P_{\bar{\nu}_{e}\rightarrow \bar{\nu}_{i}}(x)$ is the probability of
 getting a $\bar{\nu}_{i}$ at a distance $x$ form the source. Making use
 of the conservation of probability ( we will not consider oscillation
 to sterile neutrinos ) and the identity
$d\sigma^{\bar{\nu}_{\mu}}
/dT=d\sigma^{\bar{\nu}_{\tau}}/dT$ Eq. (17) can be written as

\begin{equation}
\frac{\displaystyle{d\sigma^{\nu}(E_{\nu},T)}}{\displaystyle{dT}}=
\frac{\displaystyle{d\sigma^{\bar{\nu}_{e}}(E_{\nu},T)}}{\displaystyle{dT}}
+
\left(\frac{\displaystyle{d\sigma^{\bar{\nu}_{\mu}}
(E_{\nu},T)}}{\displaystyle{dT}}
-\frac{\displaystyle{d\sigma^{\bar{\nu}_{e}}(E_{\nu},T)}}{\displaystyle{dT}}
\right)\sum_{i=\mu,\tau}P_{\bar{\nu}_{e}\rightarrow \bar{\nu}_{i}}(x)
\end{equation}

In the particular case of considering only two flavour oscillation we have:

\begin{equation}
\sum_{i=\mu,\tau}P_{\bar{\nu}_{e}\rightarrow \bar{\nu}_{i}}(x)
\rightarrow P_{\bar{\nu}_{e}\rightarrow \bar{\nu}_{\mu}}(x)=
sin^2 2\phi sin^2 \left(\frac{\displaystyle{\Delta m^2 x}}{
\displaystyle{4 E_{\nu}}}\right)
\end{equation}

\noindent
where $\phi$ is the vacuum mixing angle  and $\Delta m^2$ is the difference
 of the square of masses of the mass eigenstates $\nu_{1}$
 and $\nu_{2}$.

 From Eq. (18) it is quite evident that by measuring $d\sigma^{\bar{\nu}}
/dT$ at the kinematical configuration where $d\sigma^{\bar{\nu}_{e}}
/dT$ vanishes, the signal will be proportional to the oscillation probability
 times the $\bar{\nu}_{\mu} e^- \rightarrow \bar{\nu}_{\mu} e^-$ cross
 section thus simulating  an "appearance" experiment.

 There are some features of this appearance-like experiment which distinguish
 it from the usual appearance experiments. First, by measuring events on the
 dynamical zero we are sensitive to
 oscillations $\bar{\nu}_{e}\rightarrow \bar{\nu}_{x}$, where $\bar{\nu}_{x}$
 is any non-sterile  neutrino. The detection is not via purely charged
 current processes;
 on the contrary any signal of oscillation would be detected
 via neutral currents ($\bar{\nu}_{\mu}(\bar{\nu}_{\tau}) e^{-}
 \rightarrow \bar{\nu}_{\mu}(\bar{\nu}_{\tau}) e^{-}$). Hence there is no
 energy threshold; this experiment would use neutrinos with energies around
 0.5 MeV, thus being in principle more sensitive to low $\Delta m^2$ values
 than in the standard appearance experiments.

 In a reactor the antineutrino spectrum is continuous so in Figure (5) we have
 plotted ( dashed lines ) the curves for constant neutrino energies in the
 plane $(T,\theta)$. The solid lines represent the curves for constant
 ratio $\frac{d\sigma_{\bar{\nu}_{\mu}}}{dT}/\frac{d\sigma_{\bar{\nu}
_{e}}}{dT}$; of course they have an absolute maximum in the dynamical zero.
 From this figure it is quite evident that far from the dynamical zero
 there still remain important effects associated to its presence.

 As an illustration of the exclusion plots that one could obtain
 from the observable (18)
 we have integrated it over a typical reactor spectrum in the kinematical
 region where $\frac{d\sigma_{\bar{\nu}_{\mu}}}{dT}/
\frac{d\sigma_{\bar{\nu}_{e}}}{dT}\geq 5$ and imposed that the ratio
 $\int \frac{d\sigma^{\bar{\nu}}}{dT}
/\int\frac{d\sigma^{\bar{\nu}_{e}}}{dT}$ is less than $1.5$. With the detector
 placed at $20$ meters, the would-be exclusion plot
 we get is represented in Figure (6).
 Inside the excluded region we have inserted the by now allowed region of
 oscillations coming from atmospheric neutrino experiments
 \cite{atm}. Taking into
 account the original MUNU  proposal \cite{Bro},
 the numbers we have considered correspond
 roughly to detect a few ($\sim 10$) events per year
 if no oscillations are present
 placing an upper bound (with oscillations) around 15 events.

  We have drawn Figure (6) supposing a complete knowledge of the neutrino
 spectrum. From Figure 5 it is evident that by measuring the cross section
 at different kinematical points with the same neutrino energies we can avoid
 uncertainties from the neutrino flux. Note that for points with different
  $(\theta,T)$ but corresponding to the same energy $E_{\nu}$ the dependence
  of $\frac{d\sigma^{\nu}}{dT} (\theta,T)$ on $\Delta m^2$ and $\phi$ is
  different, so that performing ratios the dependence on the flux can be
  cancelled out without cancelling the effect.
  If this ratio is performed by integrating over a reasonable
  kinematical region we have checked that errors coming from the flux
  uncertainty can be reduced to a few percent. This result makes this
  proposal very appealing.

{\bf ACKNOWLEDGEMENTS}

\hspace{0.5cm}This paper has been supported by CICYT under Grant AEN 93-0234
 . We are
 indebted to J. Bustos, D. Koang, , M.C. Gonzalez-Garcia, F. Halzen,
L.M. Sehgal and S.K. Singh
for discussions about the topics of this paper.

\newpage
\begin{center}
{\large {\bf Figure Captions.}}
\end{center}

\begin{itemize}
     \item{{\bf Fig. 1)} Regions in the plane $(\theta,E_{\gamma})$
 where the $\bar{\nu}_{e}$
 radiative cross sections, when integrated from $\theta$ to $\theta_{max}$
 $(Q^2=0)$, satisfy that the ratio $d\sigma_{M}/d\sigma_{W}$ is 5,4,3 or 2
 times larger than the elastic ratio at the same $T$-value $(T=0.2MeV)$.
 The solid line represents the $Q^2=0$ curve. In this figure $E_{\nu}=1 MeV$.}
     \item{{\bf Fig. 2)} Ratio  of the inclusive cross sections
$d^2\sigma_{M}
/d^2\sigma_{W}$.The physical region is bounded by
 $0\leq x\leq 1$ and $0\leq \nu \leq 2E_{\nu}^2/(2E_{\nu}+mx)$; the flat region
 on the right is unphysical.}
     \item{{\bf Fig. 3)} Kinematical and dynamical zeros for the helicity
amplitudes in the plane ($E_{\nu},cos\theta$). The kinematical ones correspond
to the line $cos\theta =1$. The curves have been done for $sin^{2}\theta_{w}=
0.233$.}
     \item{{\bf Fig. 4)} Regions of dominance of weak or magnetic backward
 differential cross
 sections in the
 plane ($\mu_{\nu},E_{\nu}$)
for $\bar{\nu}_{e}$; there are three different zones divided by the
solid line. For the rest of (anti-)neutrinos there are only two regions, being
the magnetic backward cross section dominant below the corresponding line
( long-dashed for $\nu_{e}$, dashed-dotted for $\bar{\nu}_{\mu}$ and
short-dashed for $\nu_{\mu}$ ) and the opposite above the line.}
     \item{{\bf Fig. 5)} Curves for constant values of
$log(\frac{d\sigma^{\nu_{\mu}}}{dT}/ \frac{d\sigma^{\nu_{e}}}{dT})$
 ( solid lines ) and for constant $E_{\nu}$ values in MeV
 (dashed lines) in the plane $(T,\theta)$.}
     \item{{\bf Fig. 6)} Exclusion plot
 obtained by imposing that the ratio
 $\int \frac{d\sigma^{\bar{\nu}}}{dT}
/\int\frac{d\sigma^{\bar{\nu}_{e}}}{dT}$ (ratio oscillation/non-oscillation)
is less than $1.5$,
integrating the cross sections over
 a typical reactor spectrum in the kinematical
 region where $\frac{d\sigma_{\bar{\nu}_{\mu}}}{dT}/
\frac{d\sigma_{\bar{\nu}_{e}}}{dT}\geq 5$ and considering the detector is
 $20$ meters away from the reactor. The shaded zone corresponds to the allowed
 region for atmospheric $\nu_{e}\leftrightarrow\nu_{\mu}$ oscillations.}
\end{itemize}

\end{document}